**Comment on "Open is not forever: a study of vanished open access journals"**

Matan Shelomi*
Dept. of Entomology, National Taiwan University
No 27 Lane 113 Sec 4 Roosevelt Rd, Taipei 10617 Taiwan
+886 0233665588
mshelomi@ntu.edu.tw

* Corresponding Author

**Abstract**:

We comment on a recent article by Laakso et al. (arXiv:2008.11933 [cs.DL]), in which the disappearance of 176 open access journals from the Internet is noted. We argue that one reason these journals may have vanished is that they were predatory journals. The de-listing of predators from the Directory of Open Access Journals in 2014 and the abundance of predatory journals and awareness thereof in North America parsimoniously explain the temporal and geographic patterns Laakso et al. observed.

In their preprint "Open is not forever: a study of vanished open access journals," Laakso et al. [1] lament the loss of 176 open access journals from the internet. What they failed to account for in their manuscript is whether these were predatory journals, whose disappearance is not a loss, but a gain.

Far too few scientists, especially in the developing world [2], know the difference between "open access" [OA] and "predatory." It is a problem that has been heavily written about, and yet still not publicized enough; Laakso et al. make no mention of it anywhere in the manuscript. A predatory journal practices no peer review, and will instead publish quite literally everything submitted to it [3]. These journals advertise themselves as legitimate, promise quick turn-around times, claim that they are indexed in services like Web of Science or Scopus when they are not, and advertise false or bogus impact factors like "Index Copernicus." Authors who submit are required to pay article processing charges in the hundreds or even thousands of US dollars in exchange for a DOI and publication online. That would be acceptable for a legitimate journal, OA or otherwise, but since predatory journals practice no peer review, they are of no value to the scientific community or academia and any fee above zero is overpriced. A publication in a predatory journal, at least in countries and institutions aware of the problem, is considered valueless. Those who submit to predatory journals are either looking to pad their resumés, attempting to push low quality science into the record, or are naïve and do not know better [4,2]. Many of the latter authors who subsequently try to retract from the predators in order to resubmit in a legitimate journal find they are charged a massive retraction fee [5], which real journals, OA or otherwise, never charge.

Legitimate journals, OA or otherwise, are often associated with a reputable publisher or scientific association that values their past publications as highly as they do the influx of new authors, because their publication record is their reputation. Predatory journals, OA or otherwise, can be established within minutes, and do not actually care about prolonged existence because they exist solely to bilk new academics out of their grant money. If one vanishes for whatever reason, another ten can spring up in its place. Papers like Laakso et al. that extol the virtues of open access without acknowledging the criminal aspect of predators in their midst are doing the scientific community, in particular academics in the developing world, a vast disservice.

Laakso et al.'s failure to account for predatory journals lead to a methodological flaw in their analysis. The authors stated they identified vanished journals by determining "which journals have been removed from the DOAJ [Directory of Open Access Journals] by cross-checking database records from 2010–2012, 2012–2014, and 2014–2019" [1]. What the authors seem unaware of is that, prior to March 2014, the DOAJ indexed predatory journals. (Note that predatory journals, OA or otherwise, also appear on non-OA indices such as PubMed [6], so any claim that this is a criticism of DOAJ in particular or open access databases in general is unfounded.) DOAJ established criteria on that date to eliminate what they call "questionable journals," and forced all journals to re-apply for indexing [7]. Predatory journals would not have been re-indexed, and thus "vanished." This de-indexing could also have lead to vanishing from the Internet itself by leading to the failure of the predator, as fewer victims would have been led to the journal, and so their prime income source would run dry. Laakso et al's finding that most vanished journals had their last publication prior to 2015 fits this finding. If we assume that predatory journals have a mean age of 5-7 years, which Laakso et al. reported as the mean activity duration of vanished journals, then alternative explanations for the lack of 2014-2019 vanishings are that the indexes they mined to search for journals were simply more selective following 2014 and did not include such journals, that the majority of vanishings occurred during the heyday of Beall's List in the early 2010's [8,9], or that the current date of 2020 is too soon for predatory journals established after DOAJ's 2014 culling to naturally vanish.

Some interpretive aspects of the paper warrant concern. The authors found that the majority of vanished journals are from North America, which they attribute to a lack of "principles of community and OA… embedded into academic culture [there]". First, most predatory journals are registered in North America, or at least claim to be [7,10], so if the vanished journals are mostly predatory, then their North American prevalence makes sense. Second, Laakso et al.'s argument supposes that authors only publish in or are involved in editorial boards for journals based in their home country or continent, which is demonstrably false given how most journals worldwide publish in English; such that English-speaking, North American authors would have no need to consider the country or continent where the publisher is based. Finally, this supposed cultural opposition to OA is, I would argue, a straw man. Reviews of the subject have found that the main barrier for authors to publish in OA journals is not culture, but rather unwillingness to pay for APCs when other journals of equal or better reputation often charge authors nothing [11], a financial decision that would simultaneously guard such authors from publishing in predatory journals. The paper Laakso et al. cite to support their claim that "academic career progression in North America rarely provides incentives for active involvement in OA journals" [1] explicitly states this is due to "a focus on, or misunderstanding of, OA an as inherently predatory publishing practice" [12], and not due to insufficient principles of community or a disdain for sharing. The reverse situation noted in developing countries, where the quality or predatory nature of a journal has no effect on whether it is accepted for academic career evaluation, is completely independent of the OA nature of the journals: a paper is a paper [2,10,13]. Predatory journals are not a problem because they are open access; they are a problem because they are predatory. That the OA model is attractive to predators does not mean that essential activism against predators is always an attack on OA. The myth that legitimate criticism of predatory journals that happen to be OA is due to some cultural objection to "community" in science needs to go the way of predatory journals, and vanish.

The loss of a predatory journal, like the jailing of a con artist, is a cause for celebration, not lamentation. We should be happy when a predatory journal vanishes, either as a legal consequence of committing fraud or because astute authors are smart enough not to submit to them. Likewise, any authors whose papers vanished from a predatory journal that went defunct will not suffer, for either they never cared about the work beyond the line on their C.V., or because now they have an opportunity to resubmit their work to a legitimate journal, OA or otherwise, without paying usurious retraction fees. That said, legitimate OA journals vanishing would indeed be a problem, and, admittedly one can only speculate as to how many of the 176 vanished journals are actually predatory, if any. However, vanishing is equally an issue for online-only, subscription-based journals as for OA journals [14], with similar solutions to what Laakso et al. proposed, such as authors self-archiving in repositories. Open is not forever, but neither is subscription-based, and even printed material is not permanent. Assuming all problems that OA journals face are unique to them while simultaneously ignoring, downplaying, or denying the problem of predatory journals that, yes, does disproportionately affect OA publishers, helps no one but the predators.